\documentclass[preprint,12pt]{elsarticle}

\usepackage{url}

\usepackage{hyperref}
\usepackage{listings}
\usepackage{color}
\usepackage{booktabs}

\hypersetup{
    breaklinks=true,
    colorlinks=true,
    linkcolor=blue,
    citecolor=blue,
    urlcolor=cyan
}

\journal{Computer Physics Communications}

\begin{document}

\begin{frontmatter}

\title{Prototyping a ROOT-based distributed analysis workflow for HL-LHC: the CMS use case}

\author[infn,unipg]{Tommaso Tedeschi\corref{corresponding}} 
\ead{tommaso.tedeschi@pg.infn.it}

\author[cern]{Vincenzo Eduardo Padulano\corref{corresponding}} 
\ead{vincenzo.eduardo.padulano@cern.ch}

\author[infn]{Daniele Spiga} 
\ead{daniele.spiga@pg.infn.it}

\author[infn]{Diego Ciangottini} 
\ead{diego.ciangottini@pg.infn.it}

\author[infn]{Mirco Tracolli} 
\ead{m.tracolli@gmail.com}

\author[cern]{Enric Tejedor Saavedra}
\ead{enric.tejedor.saavedra@cern.ch}

\author[cern,prn]{Enrico Guiraud}
\ead{enrico.guiraud@cern.ch}

\author[lnl]{Massimo Biasotto}
\ead{massimo.biasotto@lnl.infn.it}

\cortext[corresponding]{Corresponding author}

\address[infn]{INFN Sezione di Perugia, Via A. Pascoli, 06123 Perugia, Italy}
\address[unipg]{Department of Physics and Geology, University of Perugia, Via A. Pascoli, 06123 Perugia, Italy}
\address[cern]{EP-SFT, CERN, Meyrin, 1211 Geneva, Switzerland}
\address[lnl]{INFN Laboratori Nazionali di Legnaro, Viale dell’Università 2, I-35020 Legnaro, Italy}
\address[prn]{Princeton University, Princeton, NJ 08544, USA}

\begin{abstract}
The challenges expected for the next era of the Large Hadron Collider (LHC), both in terms of storage and computing resources, provide LHC experiments with a strong motivation for evaluating ways of rethinking their computing models at many levels. Great efforts have been put into optimizing the computing resource utilization for the data analysis, which leads both to lower hardware requirements and faster turnaround for physics analyses. In this scenario, the Compact Muon Solenoid (CMS) collaboration is involved in several activities aimed at benchmarking different solutions for running High Energy Physics (HEP) analysis workflows. A promising solution is evolving software towards more user-friendly approaches featuring a declarative programming model and interactive workflows. The computing infrastructure should keep up with this trend by offering on the one side  modern interfaces, and on the other side hiding the complexity of the underlying environment, while efficiently leveraging the already deployed grid infrastructure and scaling toward opportunistic resources like public cloud or HPC centers. This article presents the first example of using the ROOT RDataFrame technology to exploit such next-generation approaches for a production-grade CMS physics analysis. A new analysis facility is created to offer users a modern interactive web interface based on JupyterLab that can leverage HTCondor-based grid resources on different geographical sites. The physics analysis is converted from a legacy iterative approach to the modern declarative approach offered by RDataFrame and distributed over multiple computing nodes. The new scenario offers not only an overall improved programming experience, but also an order of magnitude speedup increase with respect to the previous approach.
\end{abstract}

\begin{keyword}
High Energy Physics, Distributed Computing, Analysis Facility, ROOT, Dask, HTCondor
\end{keyword}

\end{frontmatter}

\section{Introduction}
\label{introduction}

Research in High Energy Physics (HEP) is characterized by complex computational challenges raised by the need for processing huge amounts of data regarding particle collisions. The largest source of such data is the Large Hadron Collider (LHC), hosted at CERN in Switzerland, which since its start has reached peaks of 1 PB/s of data generated from physics events. The machine follows a cycle of on and off periods, also called \textit{runs}. The current run, Run 3, has begun in 2022 and will last until 2025. The next run will see an upgraded hardware configuration of the machine, named High Luminosity LHC (HL-LHC)~\cite{hl-lhc}, which will start operations in 2029 and it is estimated that it will require between 50 and 100 times more computational resources than those currently used~\cite{elsen}.

Such estimate reinforces the need for developing performant software tailored to the HEP use case, something which has always been addressed in the field. Traditionally, distributed computing has been one of the main strategies to tackle processing the large physics datasets. In particular, the Worldwide Computing LHC Grid (WLCG)~\cite{wlcg} was developed in cooperation between CERN and other research institutes as a shared computing infrastructure serving all interested scientists around the world. Alongside this main distributed facility, it is common to have smaller computing clusters at the level of the single research institution.

In this context, the workflow of analysts involves developing applications that are submitted to the grid  through thousands of jobs, each processing a different set of physics events. This is enabled by the fact that the events are statistically independent, so even though they are stored in large datasets with billions of entries, they can be processed independently thus allowing for an embarrassing parallelization of the analysis. A single large-scale analysis in production run by an LHC experiment collaboration can process multiple TBs of data, involving thousands of jobs sent to the grid and spanning multiple hours or days. It also comprises two main stages, which closely resemble the MapReduce paradigm~\cite{mapreduce}: in the first stage all jobs are submitted and process different portions of the input dataset; in the second stage all the partial results need to be merged in order to produce the final desired result, which usually comes in the form of a high number of relevant statistics and plots such as histograms. This workflow becomes particularly tedious since the two stages are independently developed and deployed: users need to write separate applications to submit the initial jobs and then retrieve and merge their results.

Tackling the previously mentioned computing challenges is a matter of developing faster software as well as improving the productivity of the final users. On the one hand, increasing data processing throughput is crucial to cope with the future increasing data rates. On the other hand, ergonomic interfaces should be added to remove the lower-level programming burden from analysts. Other industries have faced similar issues and a few solutions have spawned in the data science community at large to streamline data processing pipelines with higher-level interfaces. The most widely used implementation of the MapReduce paradigm comes from the Apache software suite and in particular Apache Spark~\cite{spark} has gained wide popularity as a distributed execution engine for many types of workloads. A similar example comes from the Dask Python library~\cite{dask}, which also supports arbitrary computation graphs not strictly applicable with MapReduce.

In the same scope, users should not have to deal with building and packaging the entire software stack needed for their analysis. At the same time, each research group may need slight adjustments to their analysis algorithms and applications. It is quite often seen that many small software frameworks are developed, based on larger utility libraries used in the field. If people interested in using a specific set of libraries do not have access to exactly the same machine, creating a coherent software stack over different nodes can become a burden quite quickly. In recent years, the efforts towards improving user productivity have also started focusing on streamlining the creation of a software stack that can be easily set up and reproduced over different nodes. On the one hand, services like CVMFS~\cite{cvmfs} make it easy to ship centrally produced environments to user machines or even computing clusters. On the other hand, various institutions have begun proposing a combination of coherent and easily accessible software and hardware resources with the general label of \textit{analysis facilities}.

Although each research group may need to tweak their analysis to use specific libraries, most software environments in HEP use ROOT~\cite{root} as the tool for storing, analyzing and visualising physics data coming from the accelerator. This library defines a I/O layer and a data format through which more than 1 Exabyte of data is stored. It also offers a high-level interface to data analysis called RDataFrame~\cite{rdataframe}, which is more user-friendly with respect to other ROOT facilities and has already seen wide usage in the community.

This article highlights the recent efforts in building an infrastructure for HEP analysts that provides solutions to the aforementioned challenges. A new analysis facility is engineered on INFN (Italian national institute for nuclear physics research) resources, accessible through a web-based interface where users can develop their analysis in Python, a language that is gaining increasing popularity in the HEP community. The code can be written in a Jupyter notebook~\cite{jupyter-notebook}, through which a set of distributed resources can be accessed. As a benchmark of this new facility, a full-scale CMS analysis is ported to RDataFrame, which can take full leverage of the remote computing resources transparently while running the full application within the same notebook. Furthermore, using RDataFrame provides tangible performance improvements over the previous approach.

The document is structured as follows. Section~\ref{related-work} highlights relevant work that can be connected to some of the issues brought up so far. The main software building blocks for this work are described in Section~\ref{tools}. Section~\ref{analysis-facility-paradigm} provides more details about the concept of \textit{analysis facility} in HEP. Sections~\ref{facility-model}~and~\ref{distrdf-infn} describe more concretely the proposed new developments of this work. Section~\ref{tests} shows the results obtained in comparison with the old approach. Finally, Section~\ref{conclusions} closes the discussion and gives some perspective for future work.

\section{Related work}
\label{related-work}

The traditional programming model for HEP data analysis applications is usually based on custom loops over the events of a ROOT dataset. Each event is processed as needed, for example by filtering it out if not interesting or using it to compute new observables. A more simplified syntax was offered by TTreeFormula~\cite{ttreeformula}, a DSL within ROOT to access and process the events in a dataset. At a broader scope, different utility libraries have been developed in the field to abstract from the lower-level syntaxes, usually only of practical use in small use cases. Some examples include the nanoAOD-tools~\cite{related-nanoaod-tools} by CMS, coffea~\cite{related-coffea}, the Latinos framework~\cite{related-latinos}, Bamboo~\cite{related-bamboo} or CMGTools~\cite{related-cmg-tools}. It is worth noting that, although providing a more modern and abstract interface than what was previously available, some facilities such as RDataFrame or coffea still represent a low-level approach when compared to more experiment-specific tools such as the others cited above. This usually boils down to the fact that the latter kind of libraries usually feature functions in their API which directly refer to computations or calibrations that physicists may need to use in their daily analysis routines.

In the last few decades, most programming models in this field applied inside user code were decoupled from the job distribution, which happened by manually submitting job description files to the scheduler of a batch system such as HTCondor~\cite{htcondor} or Slurm~\cite{slurm}. In parallel, other scientific and industrial communities investigated more interactive approaches, where the programming interface could describe both the computation and the connection to the distributed resources. Apache Spark became a popular tool in a wide range of use cases, thanks to the efficient usage of resources through the MapReduce paradigm coupled with a declarative approach~\cite{related-spark-review}. Dask has gained traction in more recent years, with examples from Earth and Climate sciences~\cite{related-earth-science, related-climate} and from molecular dynamics~\cite{related-molecular-dynamics}. Although the idea of streamlining distributed HEP analysis is not a novel idea, since ROOT has been offering a parallel system for analysis called PROOF~\cite{related-proof} for many years now, the issue of making this process truly flexible and smooth for final users is yet to be solved.

The recent trend to steer towards more interactive data analysis and access to distributed resources is being picked up in the HEP community. In particular, the ingredients of HEP \textit{analysis facilities} should include modern programming models, a coherent software stack, abstracting the infrastructure for the final user and giving a full interactive analysis experience end-to-end. An example of analysis facility developed at CERN is SWAN~\cite{swan}. This is a web-based platform that all CERN users can access, giving them storage, software and computing power in the same web page. Analysts write their applications in Jupyter notebooks and also have a filesystem with storage quota at their disposal. If their notebook needs to distribute computations, a Spark cluster at CERN is made available and the user can connect to it via a GUI. Another example is given by a prototype analysis facility in the USA called coffea-casa (University of Lincoln, Nebraska)~\cite{AF_US}. This facility leverages Dask to distribute computations. It is built on top of a local Kubernetes cluster and integrates dedicated resources allocated via fairshare through an HTCondor scheduler. Similarly, another analysis facility prototype was developed at Fermilab with the label ``Elastic Analysis Facility''~\cite{us_AF_article}. The analysis facility implementation described in this work follows the same trend as the prototypes just mentioned, while aiming to involve the different geographic clusters at INFN with a novel scheduler-client connection system.

Of the available literature regarding LHC experiments analysis workflows, many were carried out with the aim of evaluating the current processes at the time (e.g. \cite{duellmann_dirk_2022_6535077}). This work instead offers a one-to-one comparison between the more modern, declarative and interactive approach offered by distributed RDataFrame and the traditional approach used in CMS data analysis.

\section{Tools}
\label{tools}

\subsection{ROOT}
\label{root}

ROOT is the most widely used software framework for storing, analysing, processing, and displaying HEP data. It has seen wide adoption at CERN and several other institutions worldwide connected with it, such as those participating in WLCG.

The framework defines a common data structure and data layout to store HEP datasets, called TTree~\cite{ttree-class}. Its layout is columnar on the disk, so that different columns can be treated independently. The ROOT I/O subsystem is able to read just a portion of a dataset, to minimize read requests to the filesystem. The minimal amount of information that can be read independently from other parts of the file is called a cluster, which corresponds to a range of entries that can belong to one or more columns. ROOT datasets can be stored and read within the local filesystem of the user machine, but very often are located in remote, distributed storage systems and can be accessed through remote protocols like HTTP or XRootD~\cite{xrootd}.

The main interface for analysing a TTree (and other data formats) within ROOT is called RDataFrame. With RDataFrame, users can focus on their analysis as a sequence of operations to be performed on the dataset, while the framework takes care of the management of the loop entries as well as low-level details such as I/O operations and parallelization, effectively creating a computation graph, that is a directed graph where each node corresponds to one of the operations to be performed on data. RDataFrame provides methods to perform the most common operations required by HEP analyses, such as \texttt{Define} to create a new column in the dataset or \texttt{Histo1D} to create a histogram out of a set of values. Other than TTree, the interface supports processing datasets stored in formats like CSV, Apache Arrow or NumPy arrays~\cite{numpy}. Users can also create an empty dataset with a certain amount of rows that can be filled through the operations in the API. This is particularly useful for benchmark and simulation scenarios.

RDataFrame has been built with parallelism in mind. In fact, it is natively able to exploit all cores of a single machine through the implicit multithreading interface available in ROOT. Moreover, the scalability of this tool is ensured by its distributed version. Indeed, RDataFrame can natively distribute physics computations on multiple nodes by splitting the analysis in tasks and executing them in a MapReduce pattern~\cite{padulano-distrdf}.

\subsection{Dask}
\label{dask}

Dask~\cite{dask} is a Python library that allows to easily parallelise existing workflows. It is mainly targeted at supporting other common Python analysis tools like Numpy~\cite{numpy} or Pandas~\cite{pandas}, but is flexible enough to accommodate any type of computation. Thus, it offers many interfaces for data processing, including machine learning and real-time analysis. In the context of this work, Dask is employed as a distributed scheduler, offering a wide set of configurations thanks to which an application can be scaled to different cluster setups like:
\begin{enumerate}
    \item Start all the remote nodes from a single machine through SSH.
    \item Leverage existing cluster deployments with Kubernetes or YARN.
    \item Connect to high performance computing resource managers that implement batch submission systems, like HTCondor, Slurm or PBS.
\end{enumerate}

Two ingredients are necessary in order to distribute computations in a Dask application. The first is the object representing the remote cluster itself, including how many resources will be assigned to it for the duration of the application. The second is an object representing the connection between the local machine and the remote cluster. This is simply called \texttt{Client} and can be used with any of the different implementations of resource managers available in Dask described above. The \texttt{Client} API allows users to asynchronously launch tasks to the remote cluster.

\subsection{XRootD}

The XRootD \cite{xrootd} framework is a C++-based suite targeting fast, low latency and scalable data access. Generically it can serve any kind of data that can fit in a hierarchical filesystem-like approach, abstracting away from the particular implementation of the data format. The core functionalities are greatly extended by a rich plugin system. It is widely used in High Energy Physics both for its remote I/O protocol and for the suite of data access tools that allow to expose the presence of large physics datasets from the storage facilities to other nodes of the Grid. The library supports caching data on one node or on a federated system of nodes, a technology that is also referred to as XCache by the community.

ROOT natively supports reading/writing files from/to remote servers via the XRootD protocol, thanks to a plugin of the TFile class. Whenever a user specifies a path that contains the \texttt{root://} prefix, that file will redirect all I/O transactions through the XRootD API.

\section{The analysis facility paradigm}
\label{analysis-facility-paradigm}

\subsection{Enhancing analysis turnaround}
\label{analysis-turnaround}
As highlighted in the introduction of this work, the challenges that LHC experiments are expected to deal with are forcing the corresponding communities to rethink their computing models, moving towards more efficient approaches. More specifically, considering the case of CMS, the introduction of NanoAOD format five years ago (an extremely reduced columnar data format, which still contains all necessary information on high-level physics objects to run an analysis \cite{NanoAOD}) pushed towards a shift in the analysis paradigm, allowing for the adoption of a quasi-interactive approach, which at the same time delivers a lower time-to-insight. Since LHC experiments share performance needs that can be tackled together, and their needs in terms of software can be made generic enough, recent R\&D was aimed at investigating industrial data-science-like approaches for the data exploration capable of efficiently exploiting the existing resources.

\subsection{Federated, distributed, heterogeneous resources}
\label{fed-dist-het-resources}
For all its computing operations, CMS exploits the previously-introduced WLCG, built up from the different resources provided by the collaboration members. The reason for this is two-fold: on the one hand, it would be difficult to concentrate all CMS operations in a single place due to the high experiment resource demands; on the other hand, in this way the experiment can exploit facilities outside of CERN, provided by member institutions from all over the world.
Computing centers were originally arranged in a strict tiered structure, with well-defined tasks for each tier: only one Tier-0, around ten Tier-1s, and a large number of Tier-2s. The last ones are responsible for providing resources for analysis, even though the differences between Tier-1s and Tier-2s have been blurring in the recent years. In addition, CMS members can also exploit available opportunistic (cloud) and specialized (HPC) resources \cite{HPC_CMS}.

\section{Developing a facility model: the strategy}
\label{facility-model}

\subsection{Implementation pillars}
\label{implementation-pillars}
The \textit{analysis facility} solution we propose is founded on a few pillars: a single central JupyterHUB~\cite{jupyterhub} for the data analysis, to which users get access interacting with a single entrypoint via CMS INDIGO-IAM, deploying their own JupyterLab~\cite{jupyterlab} instance; containerized solutions (Singularity~\cite{singularity}) to allow the user to bring their own computational environment, both in the hub and on distributed resources; the possibility to scale computations leveraging distributed computing resources from Italian Tier-2 centers, HPC or even opportunistic accessible via \path{ssh} connection; access to experiment data, obtained via XRootD protocol from the grid or local XCache instances.

\subsection{A declarative and scalable software framework}
\label{declarative-scalabele-software}
 The usage of a declarative approach is crucial since it allows the analyzer to focus on the physics itself, removing from the user scope all the boilerplate code necessary to access data, loop on events, distribute computation, and aggregate results: the time needed for the user to set up, test, tune and debug each of these steps is non-negligible, and distracts the analyzer from the physics goals. As explained in section \ref{root}, ROOT's RDataFrame interface offers a declarative solution to efficiently augment and filter data stored in NanoAOD-like file types. The users only need to specify the operations they want to run on the data, then start the analysis. The execution of the computation graph together with the retrieval of the results happens in the very same Jupyter notebook, as opposed to the various scripts that have to be run asynchronously in the legacy approach. Thus, the user can profit from the interactivity of this approach, running cells multiple times and drawing histograms directly as output of the cells. 

\subsection{A flexible and distributed computing infrastructure}
\label{flexible-distributed-infra}
In this context, the infrastructure of the facility should comply with two main characteristics: on one hand, it should be easily extensible, allowing to accommodate multiple users and different analysis needs; on the other hand, it should be capable of exploiting the very same legacy infrastructure of present Tier-2s with no additional hardware requirements. This can be achieved by leveraging Dask and its compatibility with HTCondor. In fact, the Dask-jobqueue \cite{dask-jobqueue} library allows to deploy a Dask cluster on top of an HTCondor pool. In such a way, the same resources can be used in a legacy fashion (with a batch computing approach) or quasi-interactively from a notebook after having deployed the necessary Dask cluster.

\section{Exploring distributed RDataFrame on geographic cluster at INFN}
\label{distrdf-infn}

\subsection{Infrastructure of the analysis facility}
\label{distrdf-htcondor}
\begin{figure}
\centering
\includegraphics[width=1.1\textwidth]{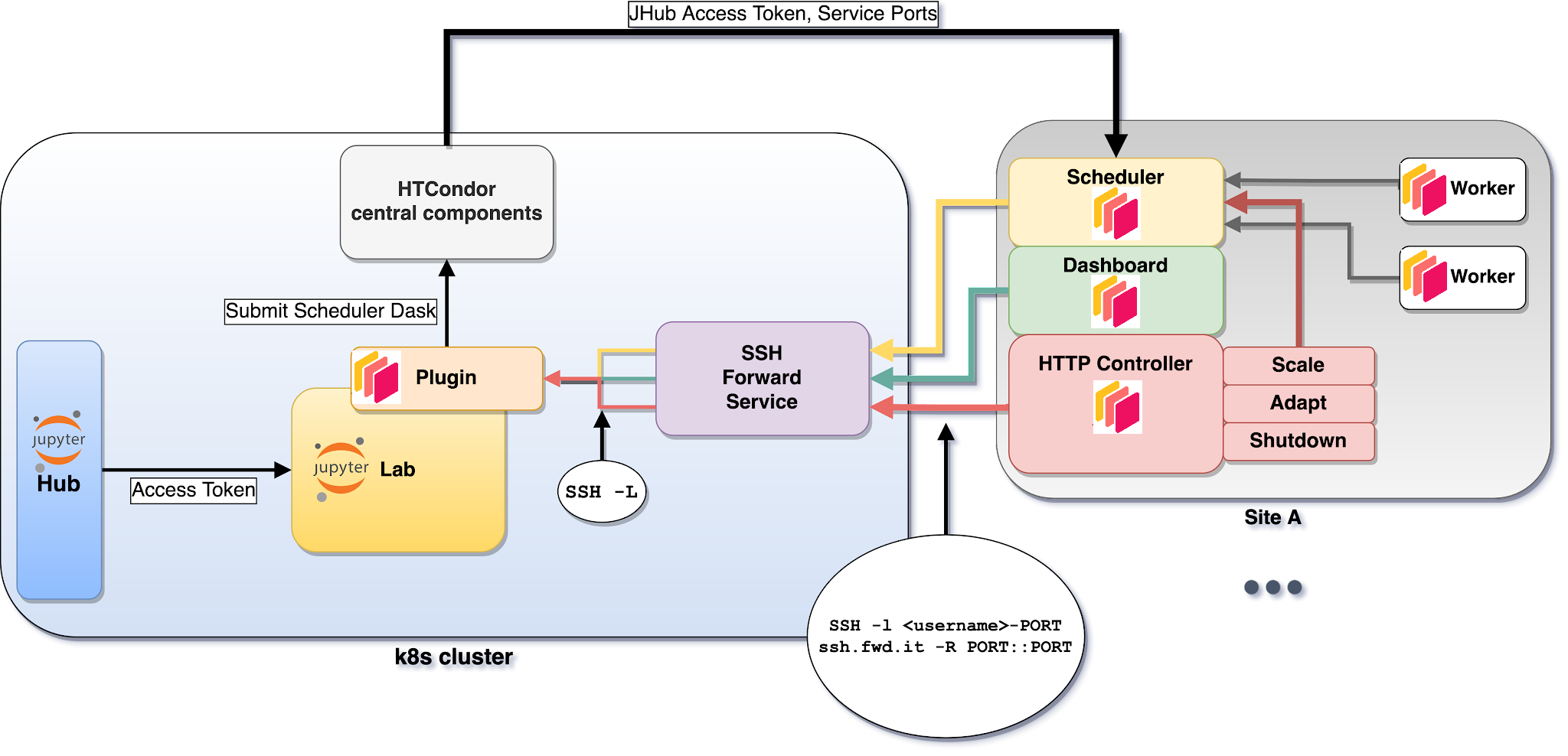}
\caption{Simple schema of INFN Analysis Facility prototype}
\label{fig:AF_flow}
\end{figure}

The ideas and practices highlighted in Sections~\ref{analysis-facility-paradigm} and \ref{facility-model} are concretely applied in the creation of an analysis facility, whose infrastructure is depicted in Figure~\ref{fig:AF_flow}. First of all, a JupyterHUB \cite{jupyterhub} instance is deployed  on the central Kubernetes cluster hosted at INFN-CNAF (Italy), as well as all HTCondor central components (Collector, Negotiator, CCB and Schedd). 
The Dask cluster deployment model presents some peculiarities enabling the shipping of a whole self-contained Dask cluster (meaning both the scheduling and executing parts) on any remote set of resources with "egress" capabilities. In fact, in the presented use case we were able to spawn a Dask cluster on a Tier-2 grid site with minimal changes from the site perspective. A full Dask cluster offloading capability is the key concept of such an R\&D that allows for the implementation of an overlay federation mechanism where an heterogeneous set of resource providers can be made available to the users in a seamless fashion, with minimal operational requirements. 
From a technical perspective, the deployment of the Dask cluster happens on top of HTCondor via the  Dask-jobqueue library, enriched with a custom-derived plugin~\cite{dask-jobqueue} (integrated with a dedicated Dask Labextension~\cite{dask-labextension}) developed to support an HTCondor pool with specific requirements and to allow the submission of the Dask Scheduler job and the interaction with it. A forwarder service is used to make Dask HTCondor jobs reachable from the JupyterLab instance via \path{ssh} connections. Finally, an HTTP controller service controls the interaction between the Dask Labextension in the JupyterLab instance and the Dask Scheduler.
The overlay system implemented through HTCondor has a key role for a fair comparison of the presented results: it allows to use the very same infrastructural setup and the very same configuration changing only the software that runs on top. 

\subsection{Prototype usage of the infrastructure}
\label{prototype-usage-infra}

The proposed infrastructure can support analysis workflows implemented in various ways, in particular it still supports the legacy batch-like approach while at the same time enabling more modern distributed workflows with RDataFrame. For the latter case, users can deploy a Dask cluster autonomously through a GUI (the Dask Labextension previously mentioned). This provides a few options, for example selecting the desired computing site or the container image for the distributed worker. Once selected, the system will submit a Dask scheduler job to the selected HTCondor pool (which in turn can exploit available Tier-2 sites, opportunistic resources, HPC facilities, etc.). Once the job is running, via the same extension, the user can scale up the cluster, submitting Dask worker jobs. As for the data access, a VOMS \cite{voms} proxy-file is needed  to be uploaded to the workers via a Dask Plugin. The user can also replicate data on the grid into a desired site via Rucio \cite{rucio1,rucio2}.

\section{Commissioning and first benchmarks with a real use case}
\label{tests}
In order to achieve the first benchmark for this infrastructure and approach, a real use case has been chosen. More specifically, the very same CMS analysis has been implemented using a legacy batch-like approach and an RDataFrame-based one, and both workflows have been run on the presented facility and compared. In this section, the details of the use case are shown, as well as the metrics used for the comparison benchmark are detailed. Finally, the results of the comparison benchmark are presented and discussed.

\subsection{The analysis use case}
\label{use-case}
This study takes into account a production-grade analysis with the CMS detector which runs, for one data taking year, over nearly 700 million Monte Carlo (MC) events (produced by the CMS Collaboration), of which more than 300k populate the final histograms considered in the statistical analysis.

\subsection{Legacy approach}
\label{legacy-approach}
The legacy approach of this analysis is based on a two-step procedure (see left column of Figure \ref{fig:porting}): a preselection step, where the original files are skimmed (using a selection based on triggers and loose requirements on objects) and corrections are computed, producing reduced flat ROOT-files; and a postselection step, in which the proper analysis is run, with the production of histograms, for each systematic variation, to be used for the final statistical analysis.
The preselection part exploits the NanoAOD-tools \cite{related-nanoaod-tools} suites, which is a collection of Python-based analysis modules orchestrated by a post-processor, developed by the CMS Collaboration.
The postselection part is run using simple PyROOT scripts, with some helper functions from NanoAOD-tools.
Both are parallelized using a simple HTCondor submission procedure. The postselection step also needs a subsequent local merging procedure to aggregate output from different jobs, as well as a local histogramming step.

\subsection{RDataFrame approach}
\label{rdf-approach}
\begin{figure}[htpb!]
\centering
\includegraphics[width=1\linewidth]{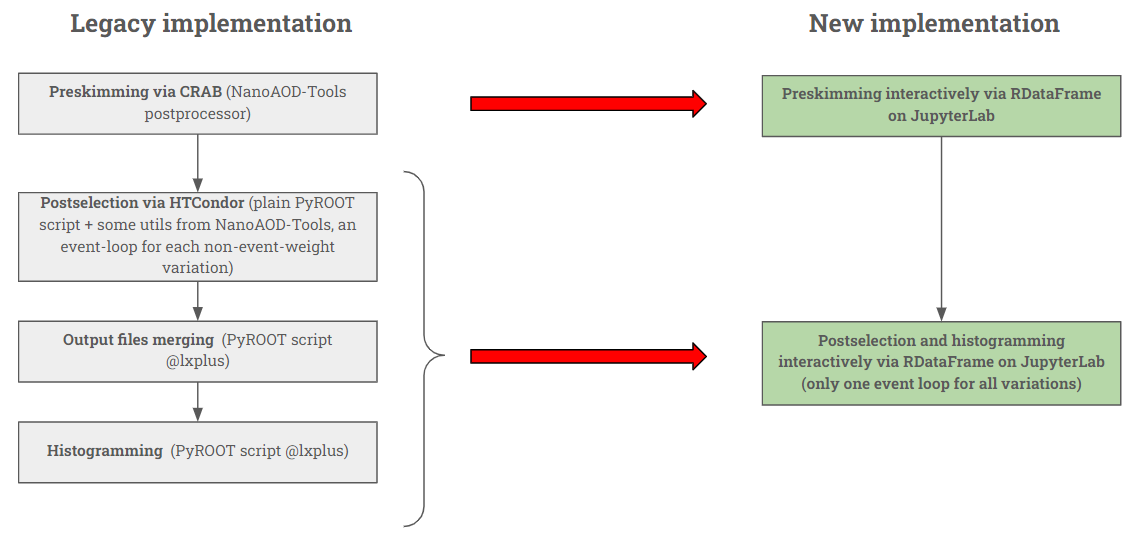}
\caption{One-to-one mapping between legacy and new approaches.}
\label{fig:porting}
\end{figure}

The new RDataFrame-based approach keeps the same workflow, in order to achieve a one-to-one mapping to the legacy approach (as depicted in Figure \ref{fig:porting}): legacy Python-based modules and functions have been translated to C++ functions that manipulate ROOT's RVec objects, also exploiting an existing solution developed inside the CMS Collaboration for jet and MET corrections \cite{JMECalculator}. This allowed to use a distributed RDataFrame approach, with Dask as backend, for both steps of analysis. The MapReduce nature of distributed RDataFrame computations makes  any merging steps unnecessary, allowing to directly produce, in a single event loop, final histograms, even including all systematic variations. In order to use Dask as a backend for the distributed execution of ROOT's RDataFrame computations from a notebook, the user needs to instantiate a Dask client and use it in the definition of a distributed RDataFrame.

\subsection{Benchmark procedure and metrics}
\label{procedure}
The testbed used for this benchmark measurement is a 3-node HTCondor pool deployed at the Tier-2 data center of LNL laboratories in Legnaro, Italy \cite{Badoer:2014fda}. Each node is a Dell R430 server with the following properties: two Intel Xeon E5-2640 v3 @2.60 GHz, 8 physical cores (16 logical) each, 128 GB of RAM, 1 TB of spinning disk storage and one ethernet controller Broadcom BCM5720, 1 Gb/s.
Each node features Telegraf \cite{telegraf} sensors, that inject metrics time-series into a dedicated influxDB \cite{influxDB} (with 1-minute granularity). In such a way, metrics values are accessible via web through an interactive dashboard. These metrics include: CPU usage percentage; amount of occupied memory; cumulated amount of data read from the network and first derivative of data read from the network (network read throughput). Moreover, monitoring scripts were also added to the single jobs in order to have complete information about the execution. More specifically, for each HTCondor job and Dask task, overall and event-loop time durations are retrieved, and CPU usage and memory occupancy (of the specific process) as functions of time are obtained leveraging psutil Python library \cite{psutil} and saved in .csv files.
Combining information from the dashboard and from the single jobs, the comparison can be made on the basis of certain metrics.
\begin{itemize}
    \item Overall execution time: time elapsed from the start of the distributed analysis execution to the end of it.  This quantifies the actual analysis time experienced by the user. A derived quantity is the overall rate (events analyzed per second), which is the ratio between the total number of events analyzed and the overall execution time.
    \item Network read: per-node information about the total amount of bytes read from the network during the execution, taken from the dashboard. This information is then summed up across all the nodes. This allows to monitor if the tool efficiently reads only what is really needed for the analysis. This measurement is crucial since the future CMS data management model (the so-called Data Lake) will strongly rely on a cache layer to distribute data to the computing centers: a minimal data read directly maps to lower requests on caches performances, as well as to higher CPU efficiency.
    \item Absolute memory occupancy: this value is directly taken from the dashboard for each node, and then it is averaged across the execution time and across all the available nodes. This metric is monitored to ensure that new approaches do not introduce any unsustainable increase in resource usage. 
    \item Job rate, which indicates the actual throughput of this approach: this value is obtained as
    \begin{equation}
        rate = \frac{\sum{\#events_i}}{\sum{t_i}}
    \end{equation}
    where $i$ is the index of $i$-th job, $t_i$ its time duration, and $\#events_i$ the number of events read by it. Depending on the way  $t_i$ is computed, one can obtain the rate quantity either including (job rate) or excluding (job event-loop rate) script initialization time. This metric is chosen since it measures the throughput of the approach with minimal dependence on job-splitting pattern or cluster size.
\end{itemize}
The benchmark comparison is done considering separately only the distributed steps of preselection and postselection: more specifically, the latter considers 3 different kinematic variables  and 30 different systematic variations (of which 8 modify the topology of the event, and thus require additional event loops in the legacy approach).

\subsection{Results}
\label{results}
The target of the benchmark is the analysis of MC  samples, simulating 2017 data-taking operating conditions, for a total of 656978035 events, divided into 1274 nanoAOD files, summing up to around 1.1 TB. The legacy approach implements, for both preselection  and postselection, one job per file, while the RDataFrame-based approach is applied using a number of partitions (i.e. Dask tasks) approximately equal to three times the number of workers in the Dask cluster. HTCondor legacy jobs require 1 CPU, while the CPU resources taken up by the Dask scheduler job and by each Dask worker job are, respectively, 4 CPUs and 1 CPU. Input and output data, for all analysis steps, are stored at the LNL laboratories Tier-2 and are accessed via the XRootD protocol.

First of all, for each scenario, the per-job (per-task) CPU usage and memory consumption were checked in order to detect pathological or wrong behaviors, like memory leaks. 

\begin{figure}[htpb!]
\centering

\includegraphics[width=0.45\linewidth]{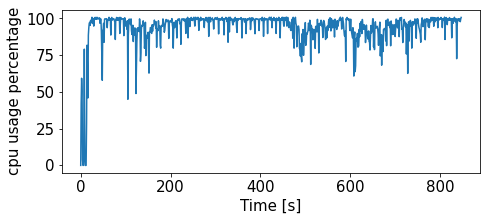}\quad
\includegraphics[width=0.45\linewidth]{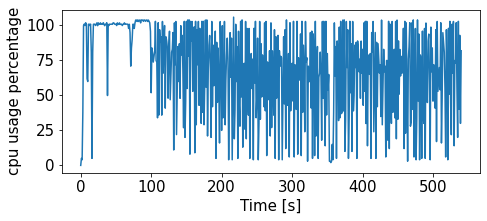}\quad
\includegraphics[width=0.45\linewidth]{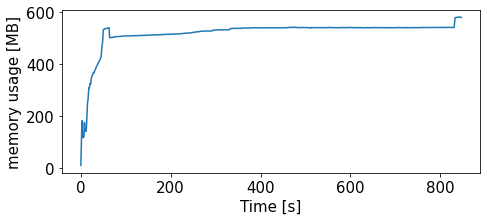}\quad
\includegraphics[width=0.45\linewidth]{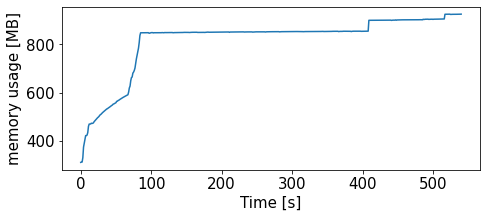}\quad
\caption{CPU usage and memory consumption for one job or task of legacy (left column) and  RDataFrame-based (right column) preselection scenario.}
\label{fig:pre_job}
\end{figure}

Figure \ref{fig:pre_job} shows CPU usage and memory occupancy, as functions of time, of one example job or task for the legacy and RDataFrame-based preselection, as they are retrieved by psutil \cite{psutil} Python library. 
As one can see, in both cases the 100\% CPU usage is reached, but the RDataFrame-based preselection task presents a noticeable oscillation in the second part of the execution, which is related to the saturation of the bandwidth (as will be highlighted in the following), while such oscillations are smaller in the other case. As for the occupied memory, an ascending behavior in the first seconds of execution is detected, which can be justified by the loading of all necessary functions and libraries, as well as by the reading of the first chunk of data. After this initial phase, the memory value remains approximately stable during the execution.

\begin{figure}[htpb!]
\centering

\includegraphics[width=0.45\linewidth]{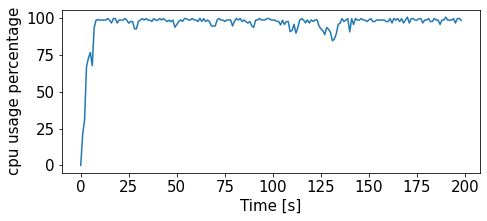}\quad
\includegraphics[width=0.45\linewidth]{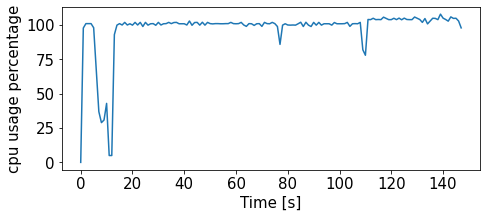}\quad
\includegraphics[width=0.45\linewidth]{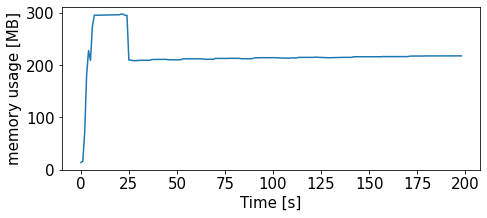}\quad
\includegraphics[width=0.45\linewidth]{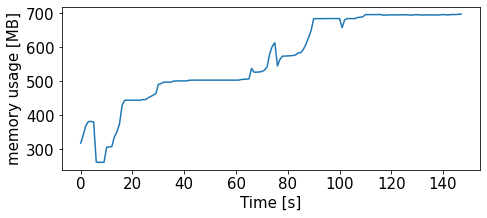}\quad
\caption{CPU usage and memory consumption for one job or task of legacy (left column) and  RDataFrame-based (right column) main postselection scenario.}
\label{fig:post2_job}
\end{figure}

Figure \ref{fig:post2_job} shows the same quantities for the postselection scenario: also here, for both approaches, the 100\% CPU usage is reached during the proper event-loop execution with no significant oscillating behavior with respect to the preselection case. Correspondingly, the stabilization of memory usage, for both approaches, happens at lower values with respect to preselection.

\begin{figure}[htpb!]
\centering

\includegraphics[width=0.4\linewidth]{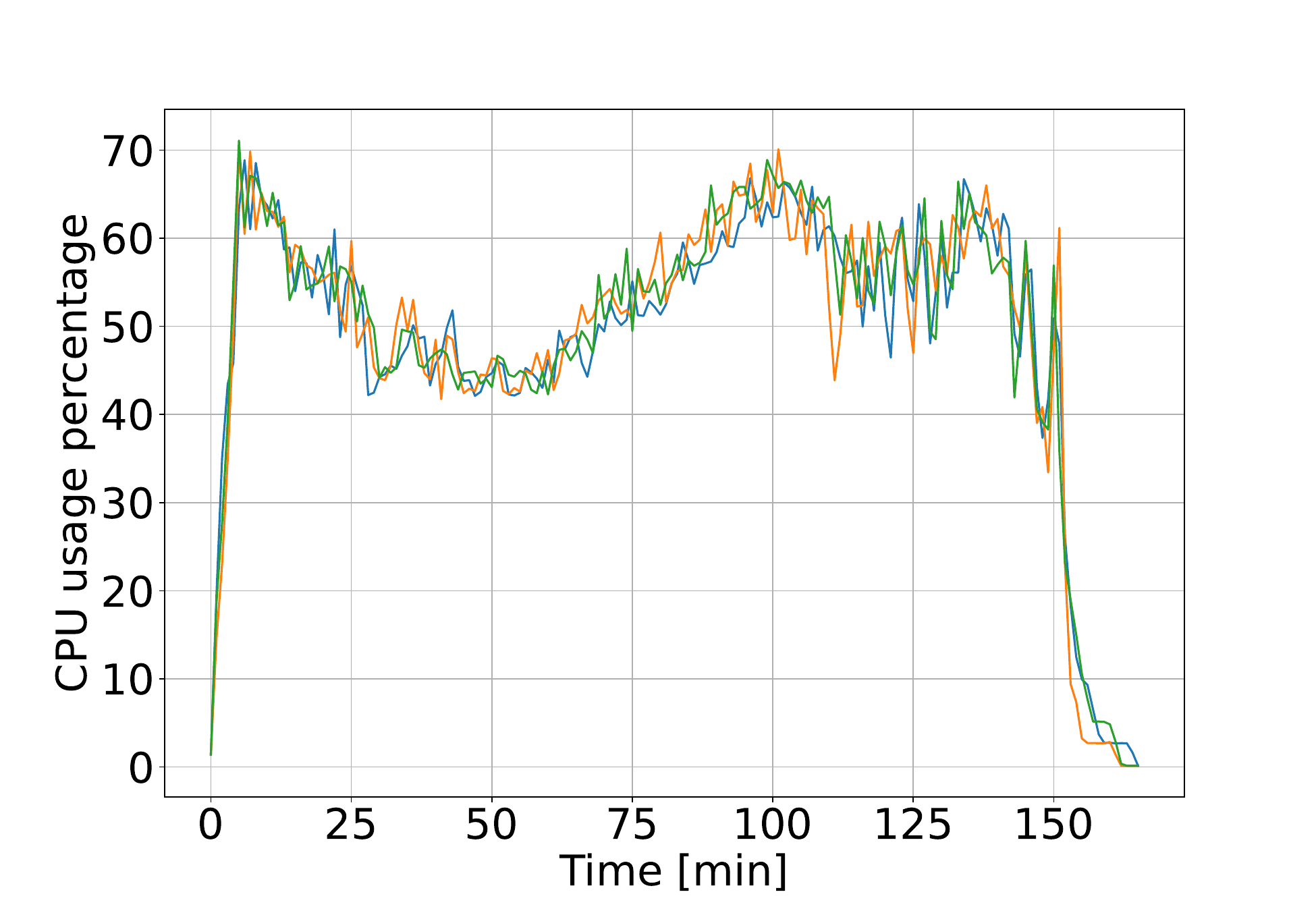}
\includegraphics[width=0.4\linewidth]{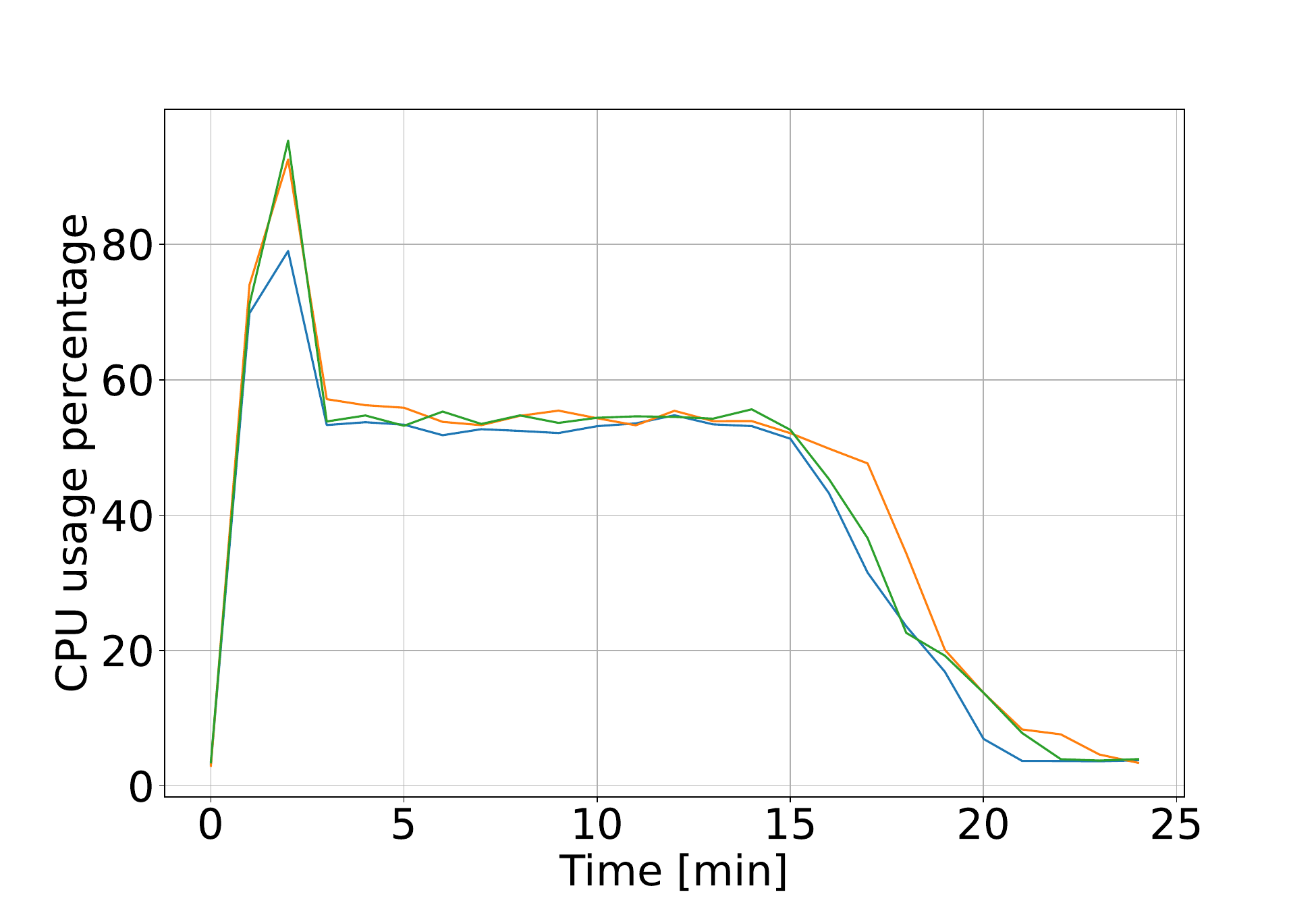}
\includegraphics[width=0.4\linewidth]{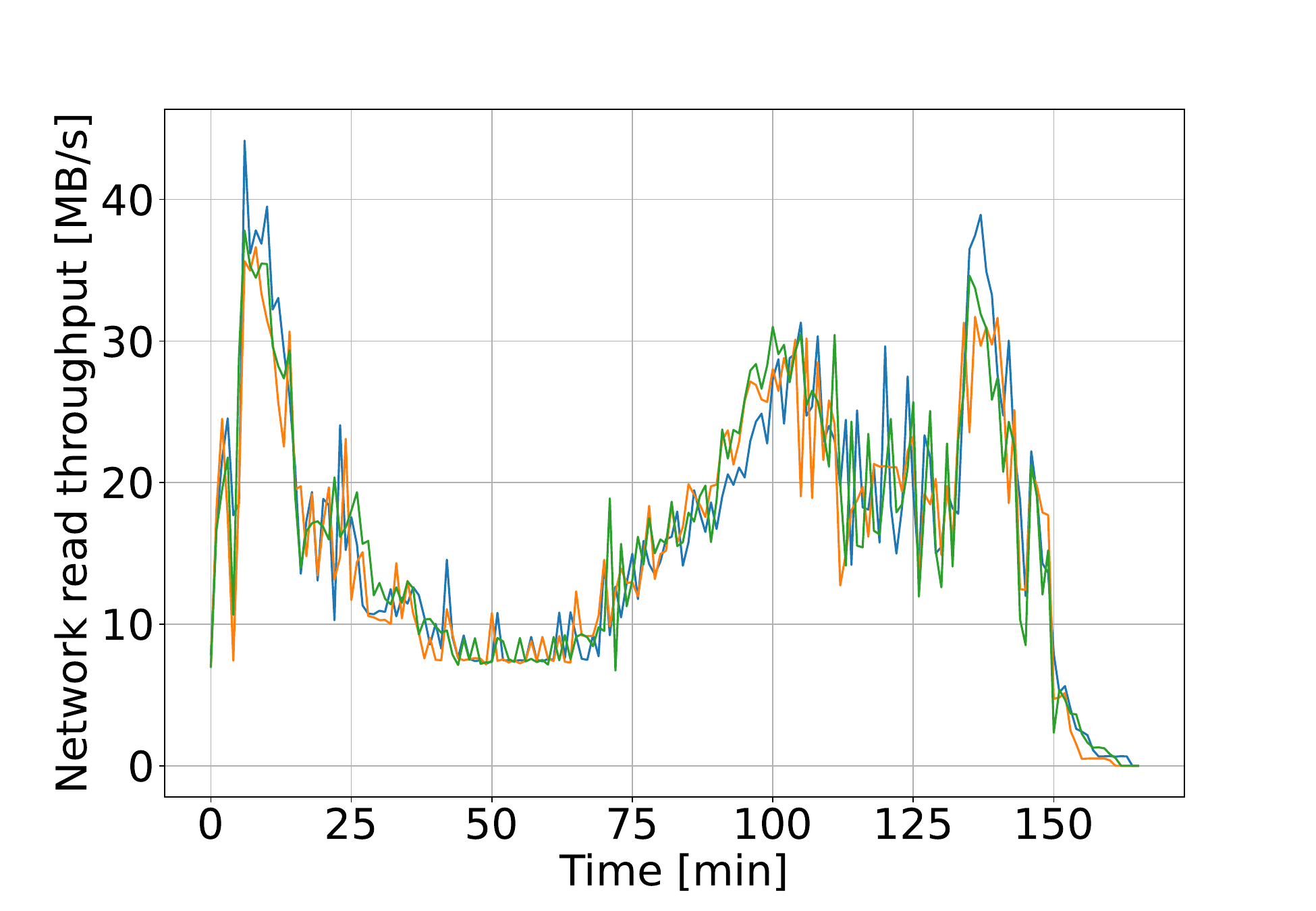}
\includegraphics[width=0.4\linewidth]{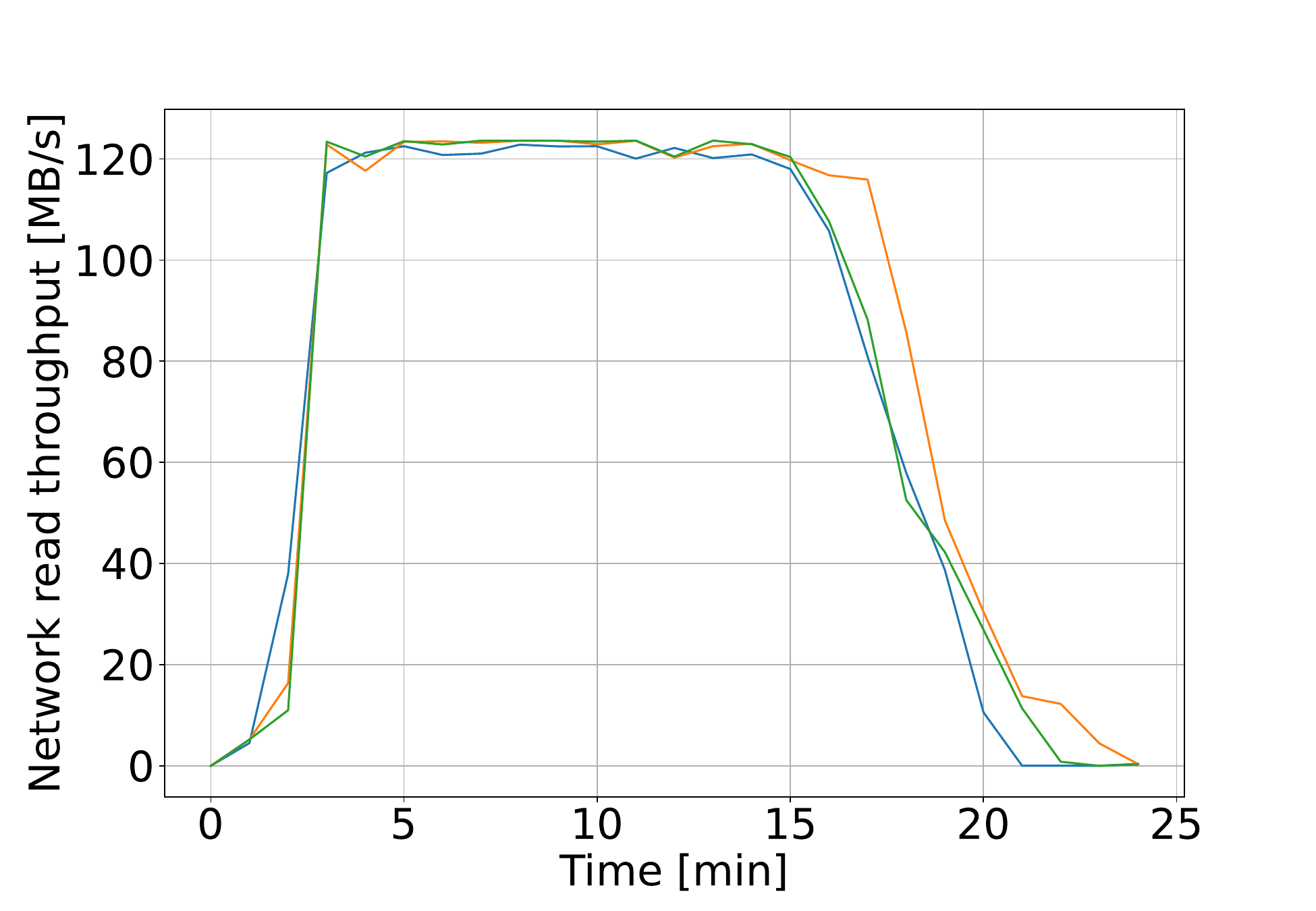}
\caption{Per-node (differently-coloured lines) CPU  usage and network read throughput for  legacy (left column) and RDataFrame-based (right column) preselection scenario, when using 96 logical CPUs (48 physical).}
\label{fig:pre_monitoring}
\end{figure}

Afterwards, the overall behaviour of the execution was checked by looking at the dashboard values. More specifically, figure \ref{fig:pre_monitoring} shows the CPU usage and the network throughput, as functions of the time of execution, reported for each one of the 3 nodes (represented by lines of different colours), in the case of legacy and RDataFrame-based preselection. As one can see, in both cases the overall execution does not reach 100\% of CPU usage: in the case of RDataFrame, this is justified by the network read throughput, which clearly reaches a plateau at around 120 MB/s, that corresponds to the nominal throughput of the network interface on the node, namely 1 Gb/s; in the case of legacy, no saturation in bandwidth is detected.

\begin{figure}[htpb!]
\centering

\includegraphics[width=0.4\linewidth]{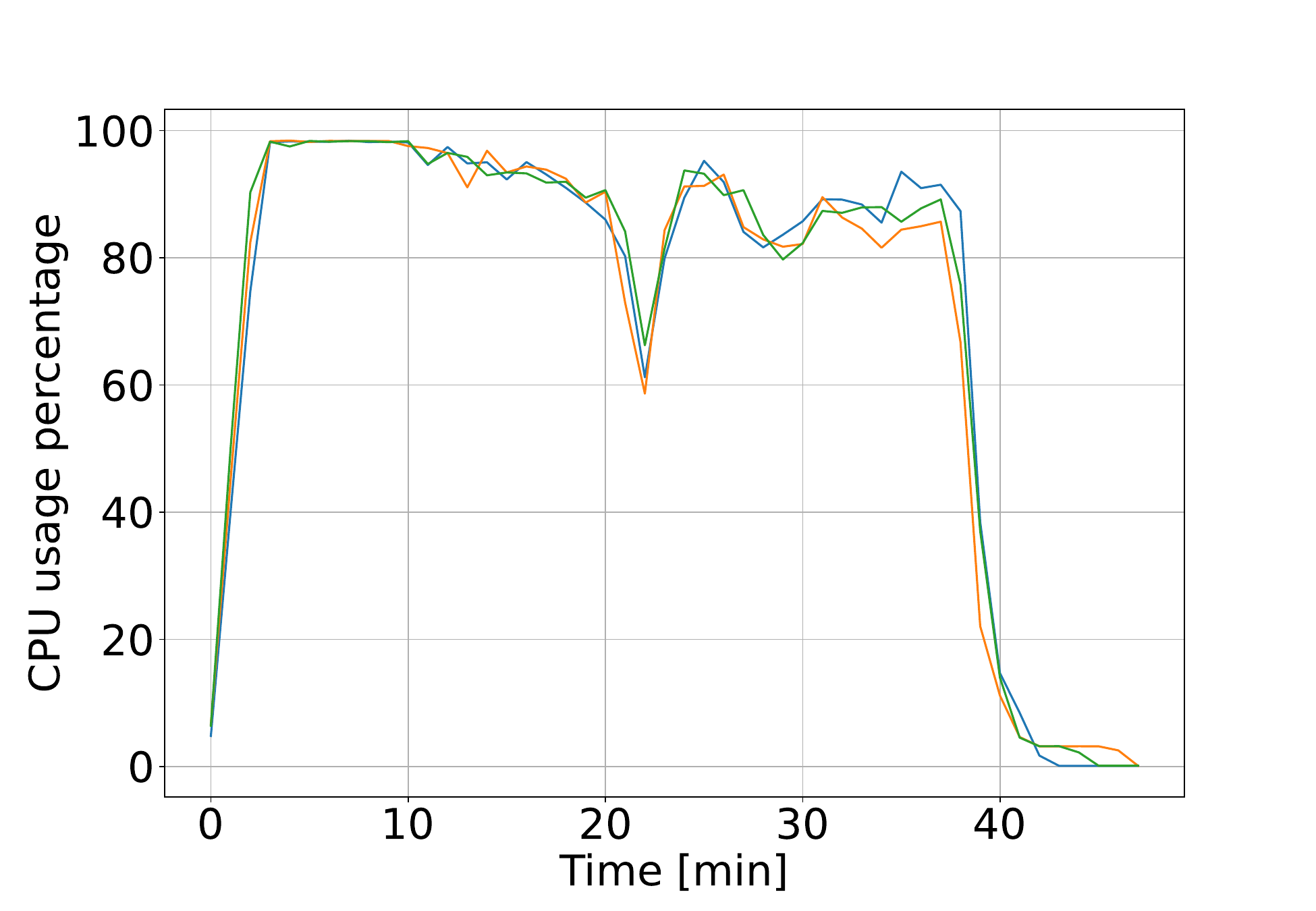}
\includegraphics[width=0.4\linewidth]{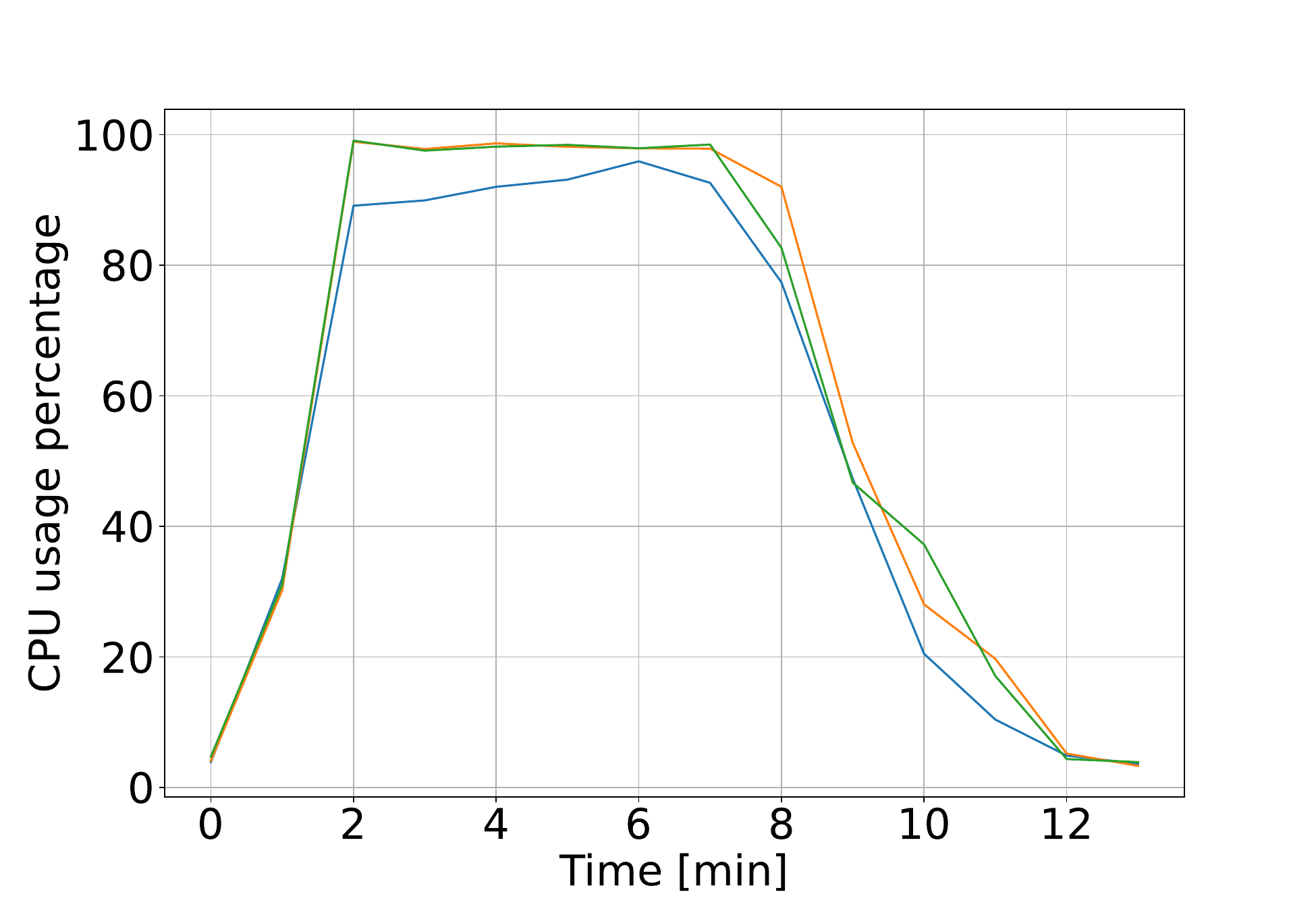}
\includegraphics[width=0.4\linewidth]{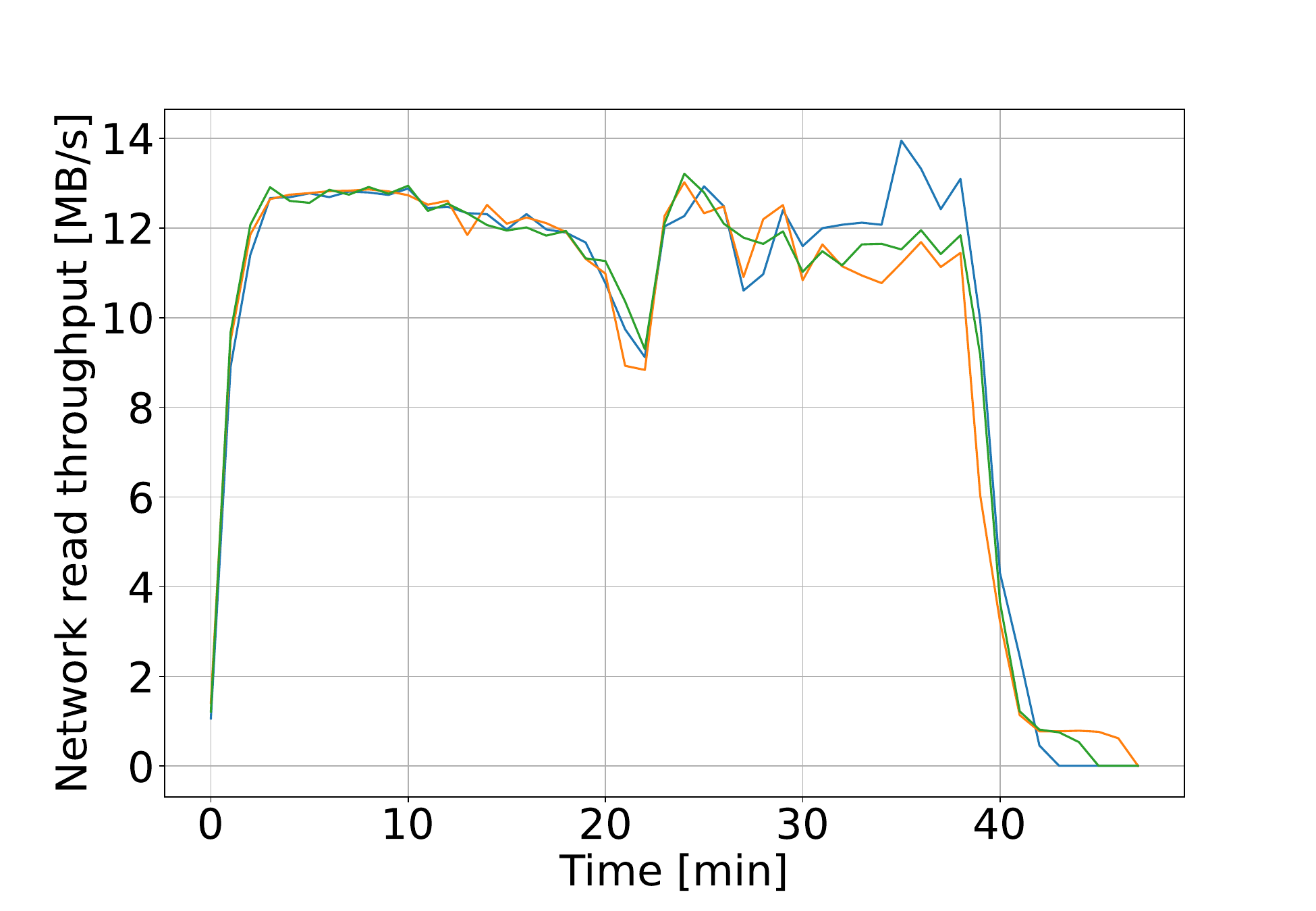}
\includegraphics[width=0.4\linewidth]{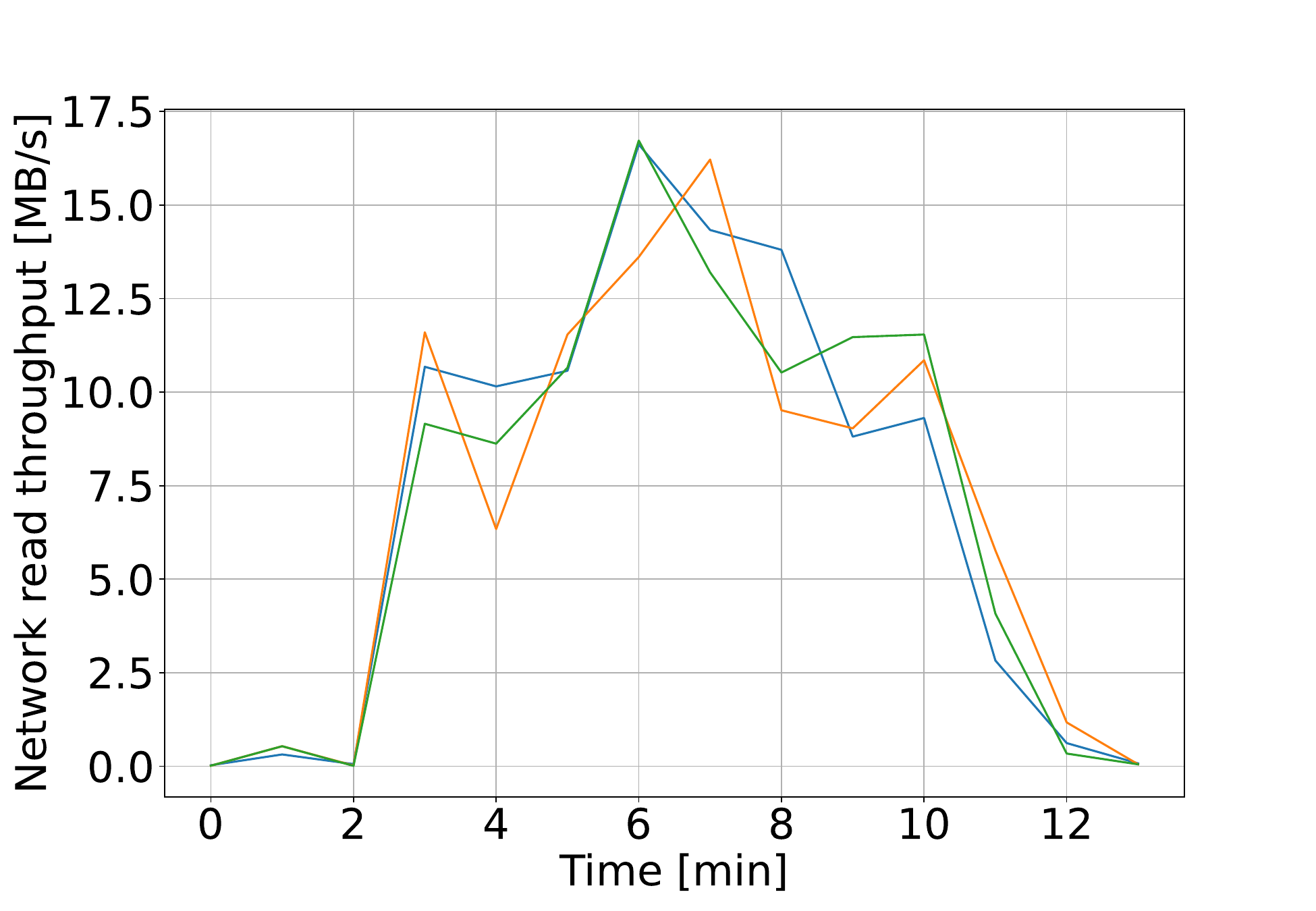}
\caption{Per-node (differently-coloured lines) CPU  usage and network read throughput for legacy (left column) and RDataFrame-based (right column) postselection scenario, when using 96 logical CPUs (48 physical).}
\label{fig:post2_monitoring}
\end{figure}

Figure \ref{fig:post2_monitoring} shows the same quantities for the postselection scenario. In this case, for both approaches, the CPU usage is nearly 100\% for most of the execution and no saturation in the bandwidth is detected.

A set of 3 measurements for each scenario and approach was performed and metrics values were recorded: then, for each metric, the average was taken as the estimated value with the maximum semi-dispersion as its error.
Results, presented separately for each scenario, are shown in table \ref{tab:post2}.

\begin{table}[htbp]
\footnotesize
\centering PRESELECTION - 96 logical CPUs (48 physical)\\
\begin{tabular}{ p{7cm}||p{2.5cm}|p{2.5cm}}
 \toprule
 \textbf{Metrics} & \textbf{Legacy} & \textbf{New}\\
Overall time [min] & $164.18 \pm 0.08$ & $21.9 \pm 0.8$  \\
Overall rate [Hz]  & $66.69$k $\pm$ $0.03$k & $500$k $\pm$ $18$k  \\
Job rate [Hz]  & $859 \pm 1$ & $7371 \pm 79$  \\
Job event-loop rate [Hz] & $951 \pm 2$ & $8148 \pm 92$  \\
Network read [GB] & $484.6 \pm 0.6$ & $353.37 \pm 0.08$  \\
Average per-node memory occupancy [GB] & $18.35 \pm 0.08$ & $29.6 \pm 0.2$  \\
 \toprule
\end{tabular}

\footnotesize
\centering POSTSELECTION - 96 logical CPUs (48 physical)\\
\begin{tabular}{ p{7cm}||p{2.5cm}|p{2.5cm}}
 \toprule
 \textbf{Metrics} & \textbf{Legacy} & \textbf{New}\\
Overall time [min] & $ 46.7 \pm 0.1$ & $11.8 \pm 0.4$  \\
Overall rate [Hz]  & $4.72$k $\pm$ $0.01$k & $18.8$k $\pm$ $0.6$k  \\
Job rate [Hz]  & $63.04 \pm 0.06$ & $303 \pm 9$  \\
Job event-loop rate [Hz] & $65.82 \pm 0.08$ & $366 \pm 13$  \\
Network read [GB] & $84.4  \pm 0.1$ & $17.56 \pm 0.06$  \\
Average per-node memory occupancy [GB] & $8.86 \pm 0.07$ & $28.3 \pm 0.3$  \\
 \toprule
\end{tabular}
\caption{Benchmark results for legacy and new approaches, for both preselection and postselection scenarios.}
\label{tab:post2}
\end{table}

\begin{figure}[htpb!]
\centering

\includegraphics[width=0.4\linewidth]{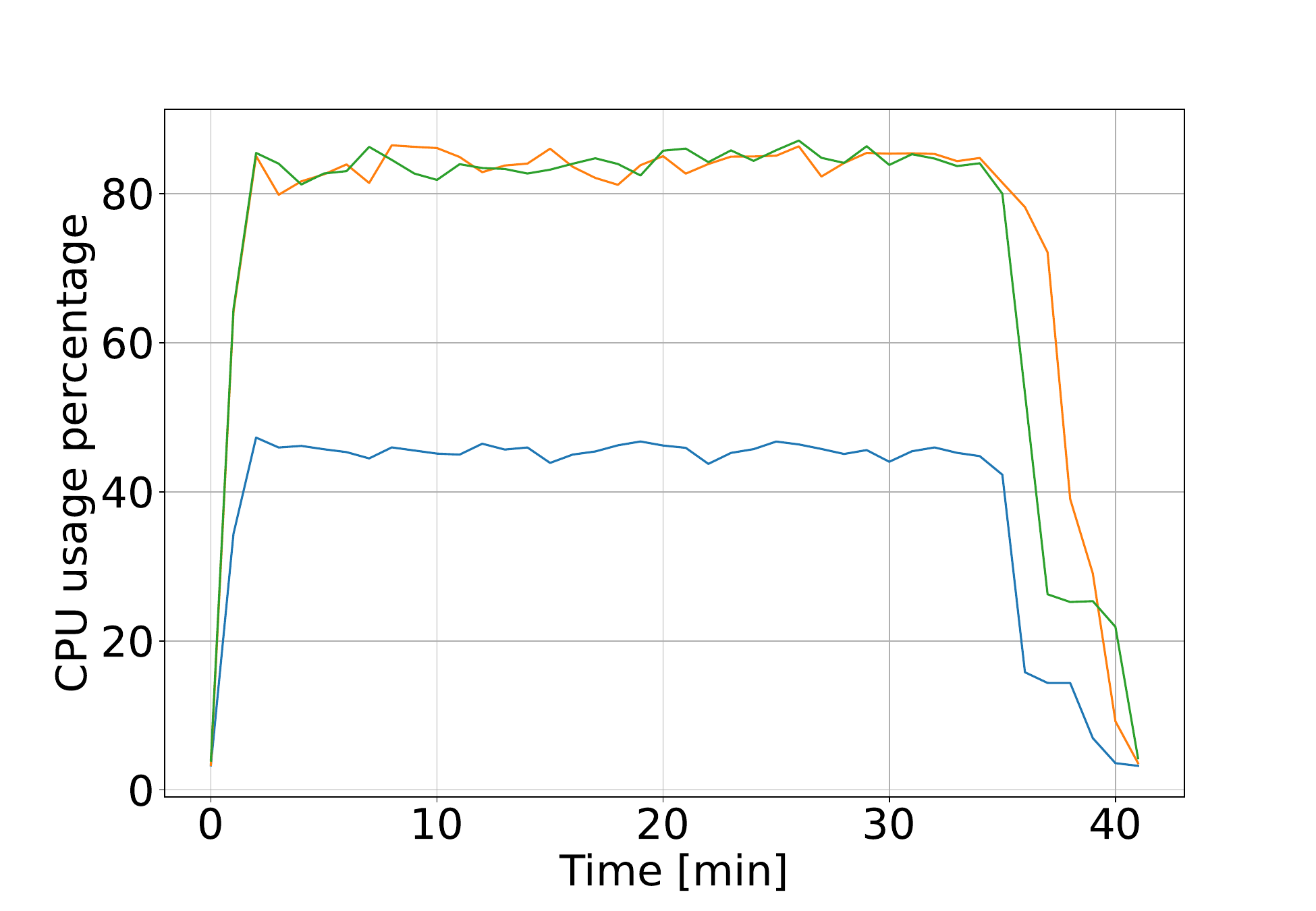}

\includegraphics[width=0.4\linewidth]{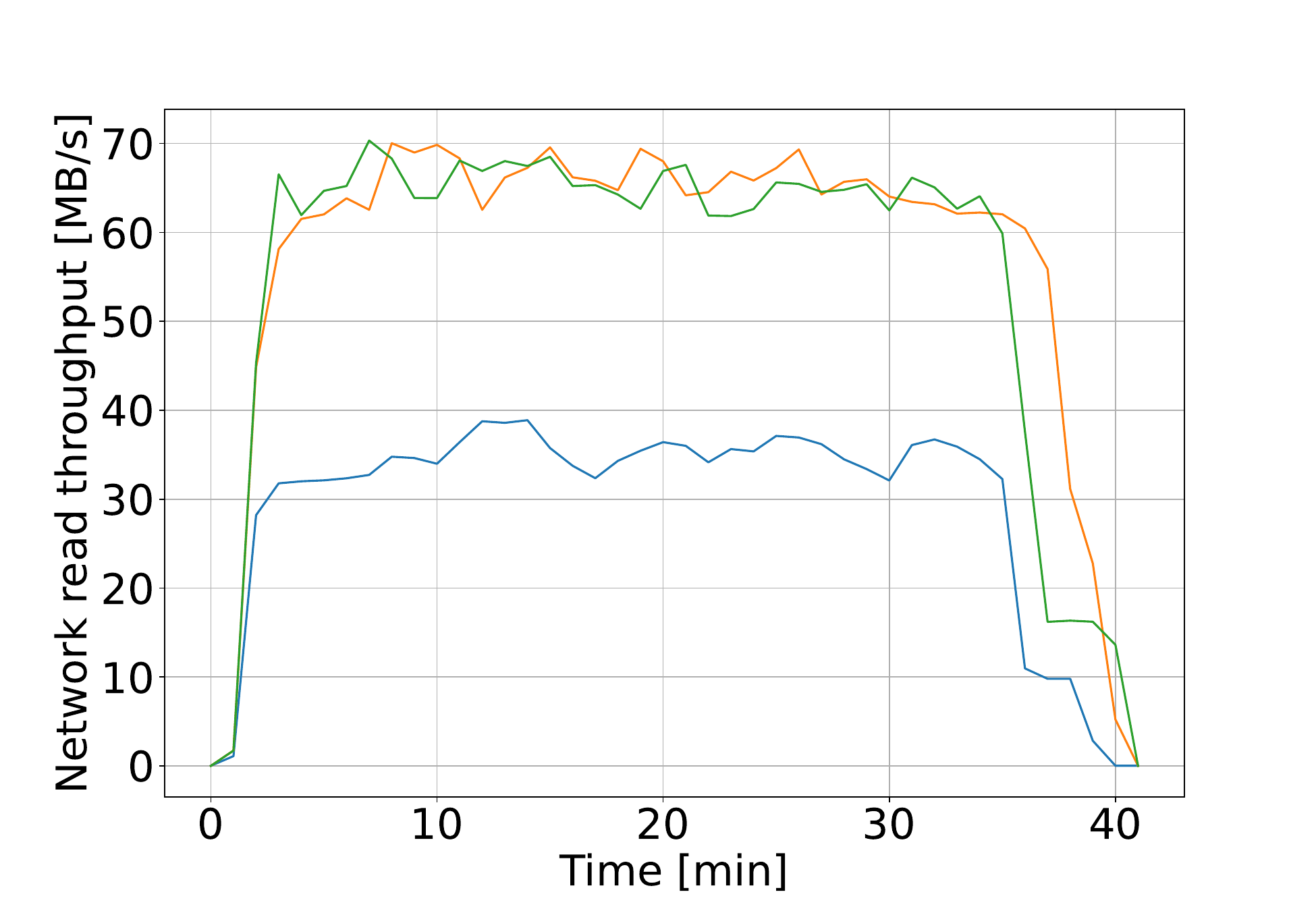}
\caption{Per-node (differently-coloured lines) CPU  usage and network read throughput for RDataFrame-based (right column) preselection scenario, when using 24 logical CPUs (24 physical).}
\label{fig:pre_monitoring_24cpu}
\end{figure}

Additionally, a check has been made in order to test if the preselection RDataFrame performance is actually limited by the bandwidth and thus the related results we present represent a lower limit. Actually, when manually lowering the number of CPUs (and thus Dask workers) per node, from the 32 logical (16 physical) CPUs to 8 logical and physical CPUs, one obtains the results shown in figure \ref{fig:pre_monitoring_24cpu}. 
As expected, keeping the same number of tasks, the RDataFrame-based CPU usage rises, reaching a plateau at over 80\% (the node with lower CPU usage is affected by the presence of the Dask scheduler), while no saturation effect is present: the network read throughput averages at 60/70 MB/s.

\subsection{Discussion}
\label{discussion}
Considering the results shown in table \ref{tab:post2}, one can conclude that the RDataFrame-based approach outperforms the legacy one in terms of time and job rate in every scenario. More specifically, considering the preselection and postselection scenarios as a whole, one can see that, moving to the new approach, a factor six speedup is achieved, corresponding to a net 84\% reduction of overall execution time. This can be justified by both the pure increase of the average job rate (around 8.6 times and 4.8 times higher for RDataFrame-based preselection and postselection, respectively) and the higher efficiency in task distribution and data read of the method itself. This means that, for an analysis similar to the one that is discussed here, a physicist can analyze, in a given time window, a factor of 6 more simulated events with respect to the old method. Furthermore, there is a reduction in network read of about 35\%. This is expected for two main reasons: on one hand, the legacy preselection step needs each job to download the full necessary nanoAOD-tools repository branch, which is worth 101 MB (summing up to 129 GB, and therefore accounting for most of the difference in network read for preselection); on the other hand, data is read 9 times in the case of legacy postselection (once for nominal values and once for each non-event-weight systematic variation), whereas, in the case of RDataFrame, just one event loop is performed. Moreover, the memory occupancy of the new approach remains bearable for such a system, since the overall node memory is 128 GB. Finally, the bandwidth-saturating behavior of RDataFrame-based preselection shows that this approach pushes the I/O capabilities of the node to the limit, as confirmed by the aforementioned additional check.

\section{Conclusions and future outlook}
The future computational challenges that High Energy Physics communities have to deal with are pushing towards intensive R\&D activities which include also the search for new efficient data analysis approaches and resource access: this translates to the adoption of declarative tools and flexible infrastructures. In this work we show the \textit{analysis facility} model implemented and deployed at INFN resources that offers a way of running CMS data analyses by accessing a single customizable JupyterLab environment and scaling up the computation on Italian geographically distributed Tier-2 resources, with the possibility to implement both batch-like approaches and distributed RDataFrame-based workflows, taking advantage of Dask and custom plugins. This infrastructure has been tested and benchmarked considering a real CMS analysis: this was implemented using  both a legacy and an RDataFrame-based approach. The two were run on the very same resources and compared on the basis of several metrics. More specifically, the modern RDataFrame approach proves to be  6 times faster, while reducing network read by more than 30 \%.
Projecting these numbers into a hypothetical HL-LHC scenario, considering the same analysis, the new approach could therefore introduce a CPU resources saving of a factor around an order of magnitude with respect to the legacy one for the same time-to-insight, also opening to the possibility of running the analysis in just 1 step, with a different impact on the resource scheduling with respect to a pure batch system (spikes of utilization of many resources instead of long-lasting jobs on few resources), and on the end-user experience. If further studies could confirm this order of magnitude of gain on a broad spectrum of different analyses (and the adoption of NanoAODs becomes even wider throughout the Collaboration), the effort in changing the CMS analysis model could be justified. More specifically, the effort would reside only on the adoption of declarative data analysis tools, since the \textit{analysis facility} concept allows to seamlessly exploit the current available WLCG infrastructure: only the software should be rewritten. On one hand, this demonstrates the strategic importance of R\&D activities for CMS, given their actual impact; on the other hand, this motivates further studies that will be funded by CMS in the future, which would possibly include the test of different tools  and the integration of legacy interfaces with the modern backends.
\label{conclusions}

\section{Acknowledgments}
The authors thank the CMS Collaboration: in particular, the CMS Offline Software and Computing groups for the technical discussions and the CMS Physics Coordination groups for the physics-related discussions, which both helped the development of this work, as well as all the other members of the Collaboration, who contributed to preparing, producing, and curating the simulated data and part of the code considered in this work.

The authors of this work are funded by the respective affiliations and this research received no specific grant from any funding agency in the public, commercial, or not-for-profit sectors.

\bibliographystyle{bibliography} 
\bibliography{bibliography}

\end{document}